\documentstyle[11pt,epsfig]{article}

\def\nue{\nu_{e}}
\def\num{\nu_{\mu}}
\def\nut{\nu_{\tau}}
\def\nmnt{$\nu_{\mu}\leftrightarrow\nu_{\tau}$~}
\def\nenm{$\nu_{e}\leftrightarrow\nu_{\mu}$~}
\def\lsim{\lower.7ex\hbox{${\buildrel < \over \sim}$}}
\def\gsim{\lower.7ex\hbox{${\buildrel > \over \sim}$}}

\begin{document}

\title{\Large \bf Results from K2K and status of T2K\footnote{Talk at International Conference on
New Trends in High-Energy Physics (Crimea2005), Yalta, Ukraine, September~10-17,~2005.}}

\author{Yuichi Oyama\footnote{E-mail address:~{\tt yuichi.oyama@kek.jp};
~~URL:~{\tt http://www-nu.kek.jp/}\~{\tt oyama}}\\ \smallskip \\
{\it High Energy Accelerator Research Organization (KEK)}}

\maketitle

\begin{center}
{\bf Abstract}\\
\smallskip
Results from the K2K experiment and status of the T2K experiment are reported.
\end{center}

\section{Results from the K2K experiment}

The K2K experiment\cite{K2K} is the first long-baseline neutrino oscillation experiment
with a distance of hundred kilometers and using an accelerator-based
neutrino beam. It started in 1999 and ended in November, 2004.
The main purpose of K2K is to confirm muon neutrino oscillation claimed
by the Super-Kamiokande experiment\cite{SKatm} using an artificial neutrino beam.
Because the details of the experiment are already described
in other articles\cite{K2K,oyaK2K},
only updated numbers and figures are presented in Table~\ref{tab:k2ksummary}.
Some comments that were not covered in my previous articles\cite{oyaK2K}
are itemized below: 

\begin{itemize}
\item After the accident of Super-Kamiokande in November, 2001, the total number of PMTs in
Super-Kamiokande was reduced to be about one half. The K2K experiment of this period
was named K2K-II. In this period, the lead-glass counters of the front detector
were replaced by a SciBar detector\cite{K2K}, a fully active fine-grained detector made
of 14848 strips of extruded scintillator read out by wavelength-shifting fibers.
There is no essential change in the oscillation analysis.

\item A possible $\nue$ appearance signal from \nenm oscillation
was also searched\cite{K2Kelec}.
From tight $e$-like event selection, only one candidate remains, where the expected
background is 1.63. The expected signal is 1 $\sim$ 2 events if the parameter region
around the CHOOZ limit\cite{CHOOZ} is assumed. The 90\% C.L. upper limit
on $\sin^{2}2\theta_{e\mu}$ (=${{1}\over{2}}\sin^{2}2\theta_{13}$)
is 0.18 for $\Delta m^{2}=2.8\times 10^{-3}{\rm eV^{2}}$.
This limit has no impact on our present knowledge on the oscillation parameters
because of the poor statistics (see Section~3.1).
\end{itemize}

\begin{table}[t!]
\caption{Summary of the K2K results. For explanations of the numbers and figures, see \cite{oyaK2K}}
\begin{center}
\begin{tabular}{ll}
\hline
\hline
~~$\bullet$ Beam period             & Jun~~4, 1999 - Jul~12, 2001 (K2K-I)\\        
                        & Jan~17, 2003 - Nov~~6, 2004 (K2K-II)\\
~~$\bullet$ Total beam time         & 442.8~days (233.7 for K2K-I + 209.1 for K2K-II)\\
~~$\bullet$ Total spill numbers     & $17.4 \times10^{6}$ spills\\
~~$\bullet$ Total POT for analysis  & $92.2 \times 10^{18}$\\
\hline
~~$\bullet$ Total event (data/expectation)    & 112~~~~/~~~~~155.9\hbox{${{+13.6}\atop{-15.6}}$} \\
~~~~~~~~~~~~Single ring events                & ~67~~~~/~~~~~~99.0\\
~~~~~~~~~~~~~~~~~$\mu$-like                   & ~58~~~~/~~~~~~90.8\\
~~~~~~~~~~~~~~~~~$e$-like                     & ~~9~~~~/~~~~~~~8.2\\
~~~~~~~~~~~~~~~~~~~~~~(tight e-like cut)      & ~(1)~~~/~~~~~~(1.63)\\
~~~~~~~~~~~~Multi ring events                 & ~45~~~~/~~~~~~56.8\\
\hline
~~$\bullet$ Null oscillation probability    & 0.003\%\\
~~$\bullet$ Best fit parameters in physical region     & ($\Delta m^{2}$,$\sin^{2}2\theta$)
            =($2.76\times 10^{-3}{\rm eV^{2}}$,1.0)\\
~~$\bullet$ 90\% C.L. $\Delta m^{2}$ for $\sin^{2}2\theta=1$ & $(1.88 \sim 3.48)
\times 10^{-3} {\rm eV^{2}}$\\
\hline
~~$\bullet$ Reconstructed neutrino energy spectrum                & 
~~$\bullet$ Allowed region in $\Delta m^{2}$-$\sin^{2}2\theta$ plane\\
~~~~\includegraphics[height=6.0cm]{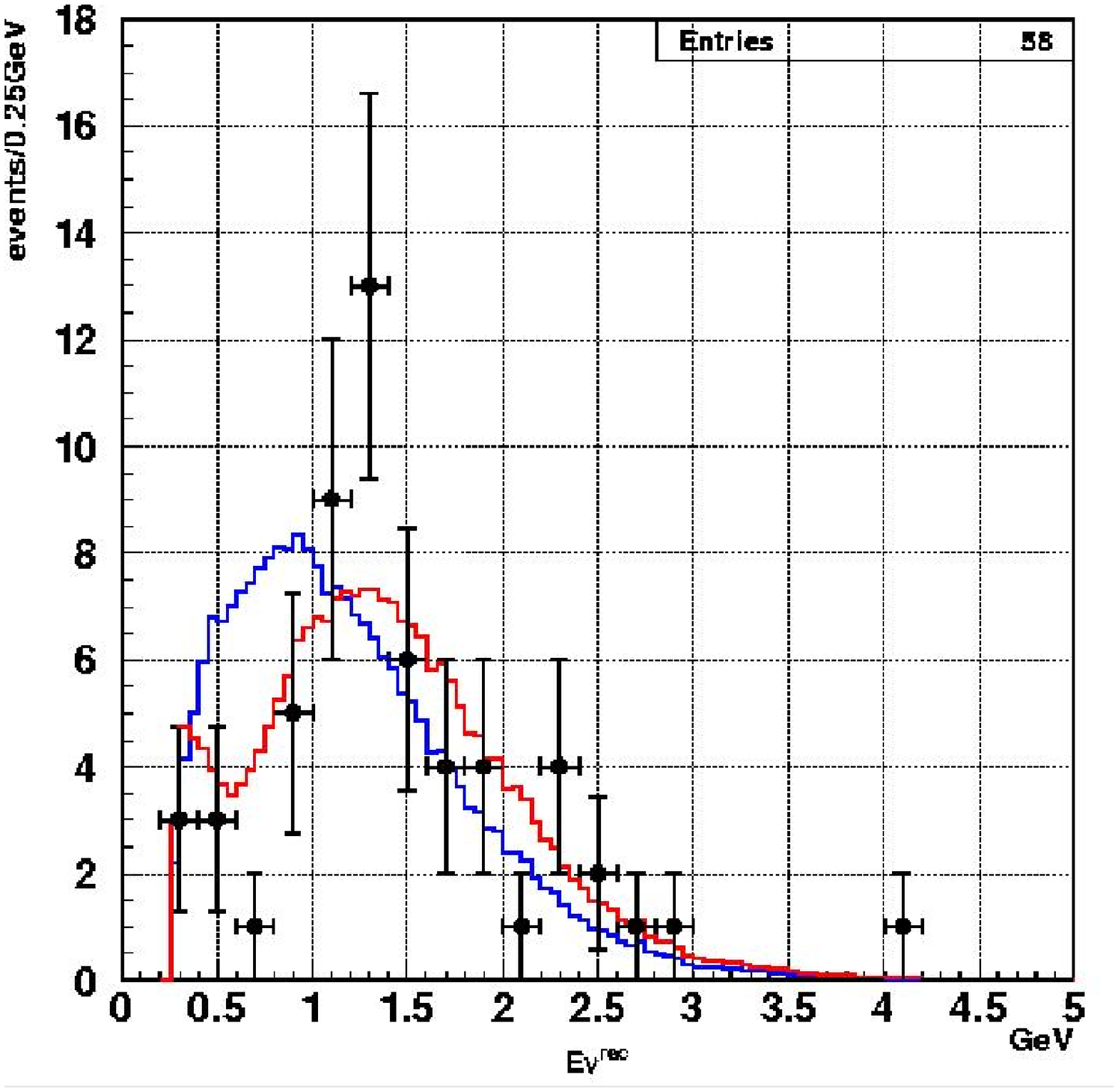}    &
~~~~\includegraphics[height=6.0cm]{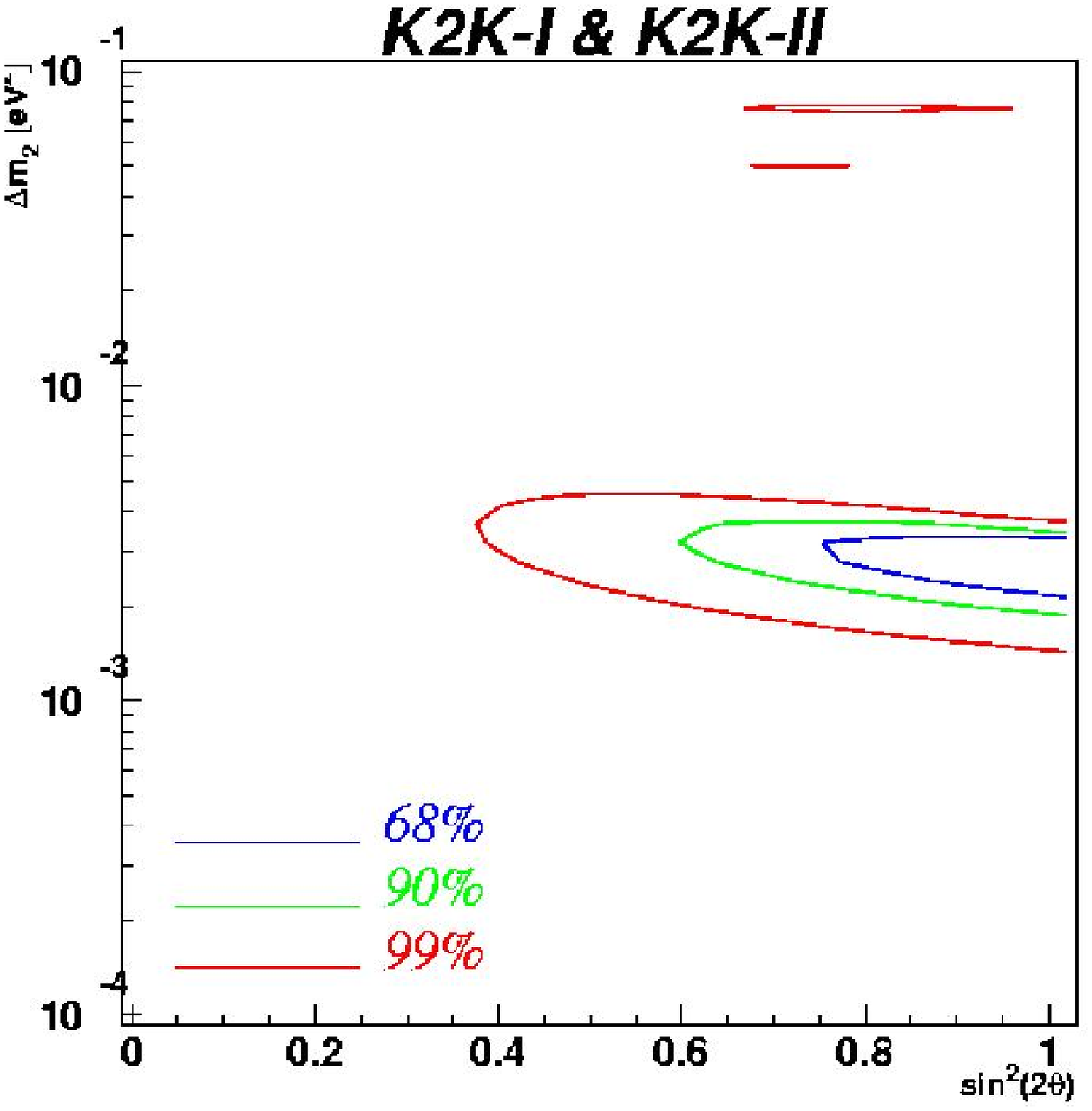}\\
\hline
\hline
\end{tabular}
\end{center}
\label{tab:k2ksummary}
\end{table}

\section{The T2K project: beamline and detectors}

T2K (Tokai to Kamioka)\cite{T2KLOI} is the next long-baseline neutrino-oscillation experiment
in Japan. A high-intensity neutrino beam from the J-PARC 50GeV Proton Synchrotron
in JAEA, Tokai is shot toward Super-Kamiokande, 295~km away.
Since all of these facilities, except Super-Kamiokande, are not familiar in the high-energy
physics society, they are summarized in Table~\ref{tab:t2kword}. A bird's eye view illustration of
the entire J-PARC is shown in Fig.~1.

\begin{table}[b!]
\label{tab:t2kword}
\caption{Summary of the facilities and nicknames related to the T2K experiment.}
\begin{center}
\begin{tabular}{ll}
\hline
\hline
J-PARC    & \underline{J}apan \underline{P}roton \underline{A}ccelerator
             \underline{R}esearch \underline{C}omplex. The name of the entire project.\\
          & It includes high energy physics, nuclear physics, life science, material science \\
            & and nuclear technology. The Accelerators consist of 400MeV Linac, \\
            & 3~GeV Proton Synchrotron and 50GeV proton Synchrotron.\\
JAEA      & \underline{J}apan \underline{A}tomic \underline{E}nergy \underline{A}gency.
              Host institute of J-PARC. KEK is the second host institute.\\
~~          & Renamed from JAERI (\underline{J}apan \underline{A}tomic \underline{E}nergy \underline{R}esearch
              \underline{I}nstitute) on October 1, 2005.\\ 
Tokai     & Name of the village where JAEA is located. About 110km north-east from Tokyo\\
JHF       & \underline{J}apan \underline{H}adron \underline{F}acility.
              Name of the 50GeV Proton Synchrotron project.\\
T2K       & \underline{T}okai \underline{to} \underline{K}amioka. Name of the long-baseline
              neutrino-oscillation experiment.\\ 
            &It was also called JHF-$\nu$ or J-PARC~$\nu$\\
\hline
\hline
\end{tabular}
\end{center}
\vskip 0.5cm
\end{table}

\begin{figure}[t!]
\center{\includegraphics[height=7.0cm]{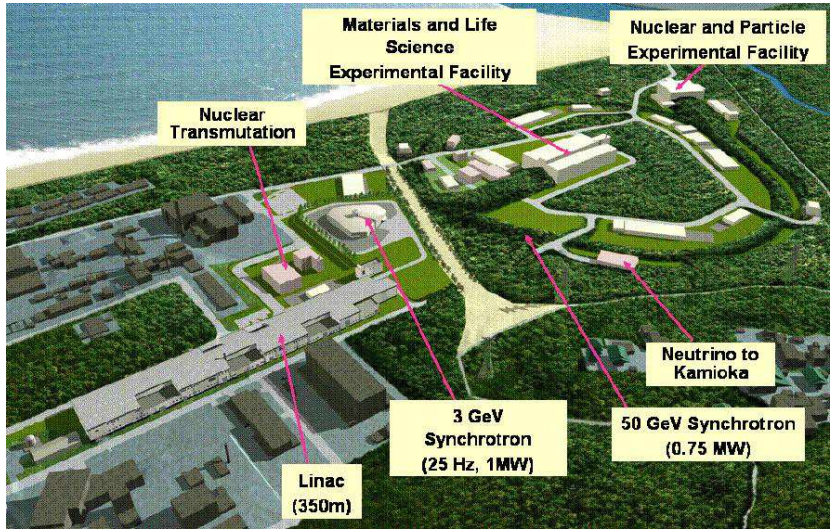}}
\center{
\caption{
Bird's eye view of the entire J-PARC project.
}}
\label{fig:birdview}
\end{figure}

J-PARC has been under construction since 2001. The T2K experiment was officially
approved in December, 2003, except for the 2-km detector discussed below.
Construction of the neutrino beamline started in April, 2004.
The experiment will start in early 2009.

Schematic configurations of the T2K beamline and detectors are illustrated
in Fig.~2.
They consist of the proton beamline, target station, decay volume,
beam dump, muon monitors at
140~m downstream from the target, first front detectors at 280~m,
second front detectors at 2~km, and Super-Kamiokande as a far detector
at 295~km.
The main differences from the K2K experiment are: (1)~very high beam intensity,
(2)~off-axis beam and (3)~2~km detector. In the following, these three main differences
are focused on one by one.

\subsection{High-intensity proton beam}

The most significant upgrade from the K2K experiment is the beam intensity.
The beam power of T2K in the first stage is 0.75~MW. It is more
than 2 orders of magnitude larger than that of K2K. A further upgrade of the beam intensity
after several years of 0.75~MW operation is also under consideration.
Comparisons of the beam parameters between the K2K, T2K and T2K upgrade
are summarized in Table~\ref{tab:beamintensity}

\begin{figure}[b!]
\center{\includegraphics[height=4.0cm]{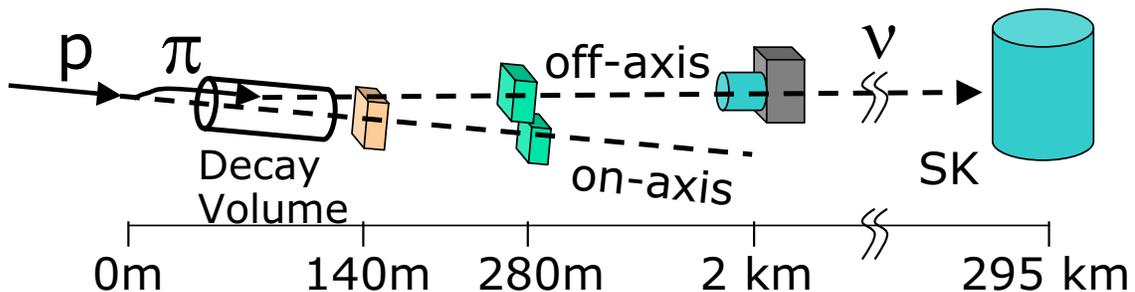}}
\center{
\caption{
Schematic configuration of the T2K beamline and detectors.
}}
\label{fig:schematic}
\end{figure}

The number of neutrino events in T2K is about 110-times larger than that of K2K.
This high-statistics observation enables searches for
unknown oscillation channels as well as precise determinations of the
oscillation parameters in \nmnt . 

\subsection{Off-axis beam}

\begin{table}[t!]
\caption{
Comparisons of the beam parameters between K2K and T2K.
For the neutrino events in Super-Kamiokande,
22.5kton is the fiducial mass and one year is taken as 100~days.
The absence of neutrino oscillation is assumed, and the number for K2K was
calculated from 155.9events/442.8days~(see Table 1).
For T2K, a 2$^{\circ}$ off-axis angle is assumed.
Some of the parameters in the T2K upgrade are still being designed.
}
\smallskip
\begin{center}
\begin{tabular}{lccc}
\hline
\hline
                          &     K2K    &      T2K      &  T2K upgrade \\
\hline
Proton Energy (GeV)       &    12      & 50      &  50   \\
Beam power    (kW)        & 5.2        & 750     &  4000  \\
Proton per second         &  $3 \times 10^{12}$   &  $1 \times 10^{14}$ & $5 \times 10^{14}$\\
Accelerator cycle (sec)    &  2.2   &  3.64 & \\
Beam duration ($\mu$sec)   &  1.2   &  4.2 & \\
Neutrino events in SK (/22.5kton/year)   &  35   &  3900 & \\
\hline
\hline
\end{tabular}
\end{center}
\label{tab:beamintensity}
\end{table}

Off-axis beam\cite{offaxis} means that
the center of the beam direction is adjusted to be
$2^{\circ} \sim 3^{\circ}$ off from the Super-Kamiokande direction,
as shown in Fig.~2.
Although the neutrino beam intensity at Super-Kamiokande is lower than that of the
beam center (on-axis) direction,
the peak energy is low and high-energy neutrinos are strongly suppressed.
The neutrino energy spectra for several off-axis angles are
shown in Fig.~3.
An off-axis beam is favored because a neutrino energy lower than
$\sim$1~GeV is preferable in the T2K experiment for the following three reasons:

\begin{figure}[b!]
\center{\includegraphics[height=6.0cm]{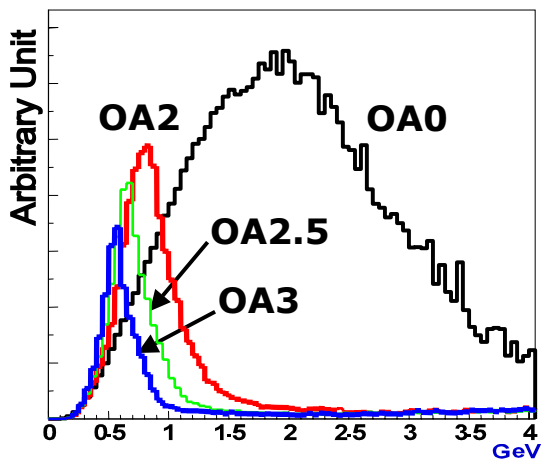}}
\center{
\caption{
Neutrino energy spectrum for several off-axis angles.
}}
\label{fig:offaxis}
\end{figure}

\begin{figure}[t!]
\center{\includegraphics[height=7.0cm]{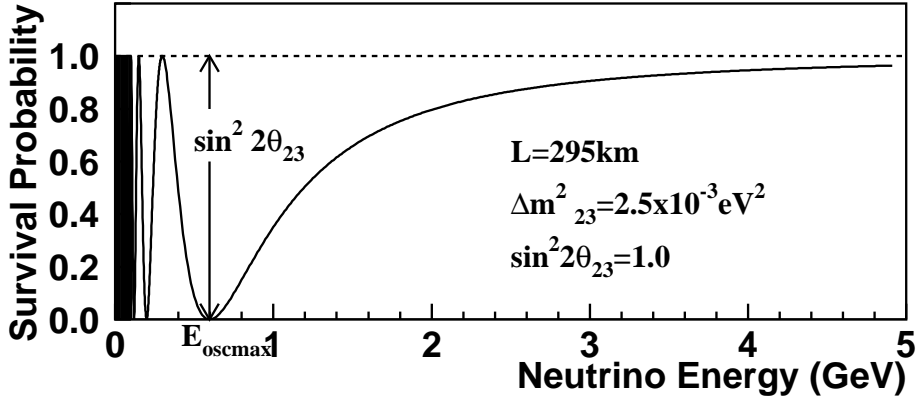}}
\center{
\caption{
Survival probability of muon neutrinos as a function of the neutrino energy.
The neutrino travel distance is 295~km and oscillation parameters are assumed
to be $(\Delta m^{2}_{23}, \sin^{2}2\theta_{23})=(2.5\times 10^{-3}{\rm eV^{2}},1.0)$.
}}
\label{fig:oscprob}
\end{figure}

The first reason is the neutrino oscillation probability.
The probability that $\nu_{\mu}$ remains as $\nu_{\mu}$
is written as 
\begin{equation}
P(\nu_{\mu} \rightarrow \nu_{\mu}) = 
1 - \sin^{2}2\theta_{23} \sin^{2}\biggl({{1.27 \Delta m^{2}_{23}L}\over{E_{\nu}}}\biggr), 
\end{equation}
where $L$ is the neutrino travel distance ($L$ = 295km) and
$E_{\nu}$ is the neutrino energy.
$P(\nu_{\mu} \rightarrow \nu_{\mu})$ as a function of $E_{\nu}$ is also shown
in Fig.~4.
The neutrino energy, which corresponds to the first oscillation maximum, $E_{oscmax}$,
is obtained from
\begin{equation}
{{1.27 \Delta m^{2}_{23}L}\over{E_{oscmax}}}={{\pi}\over{2}}
\end{equation}
or
\begin{equation}
E_{oscmax}={{2.54 \Delta m^{2} L}\over{\pi}}.
\end{equation}
$E_{oscmax}$ = 0.596GeV for $\Delta m^{2}=2.5\times 10^{-3}{\rm eV}^{2}$ and
$E_{oscmax}$ = 0.834GeV for $\Delta m^{2}=3.5\times 10^{-3}{\rm eV}^{2}$.  
If the neutrino beam energy agrees with $E_{oscmax}$,
the oscillation occurs effectively, and a study of neutrino
oscillation is also efficient. 

The second reason is the fraction of the charged-current quasi-elastic scattering
(CCQE) in the neutrino interactions.
As reported in K2K\cite{K2K}, the neutrino energy spectrum is obtained
from CCQE interactions,
\begin{equation}
\nu_{\mu} + n \rightarrow \mu^{-} + p,
\end{equation}
calculated by simple 2-body kinematics
\begin{equation}
E_{\nu}={{m_{N} E_{\mu}-m^{2}_{\mu}/2}\over{m_{N}-E_{\mu}+p_{\mu} \cos\theta_{\mu}}},
\end{equation}
where $m_{N}$ and $m_{\mu}$ are the masses of the nucleon and the muon, respectively, and
$E_{\mu}=\sqrt{p_{\mu}^{2}+m_{\mu}^{2}}$.
Other neutrino interactions cannot be used for this purpose because of complex
kinematics. Moreover they constitute serious background when CCQE
events are selected.
The neutrino cross section and the contribution of CCQE interactions are shown is
Fig.~5. Although neutrinos of energy larger than $\sim$1~GeV have larger total cross
sections, most of the interactions are non-CCQE and disturb
the determination of neutrino energy spectrum.

\begin{figure}[t!]
\center{\includegraphics[height=8.0cm]{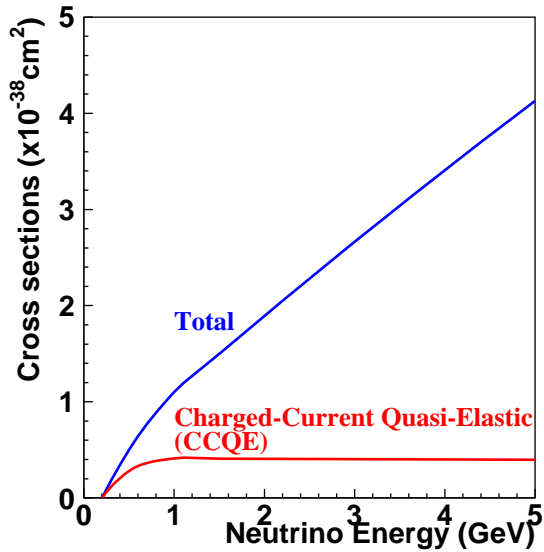}}
\center{
\caption{
Neutrino cross section and the fraction of CCQE as a function of the neutrino energy.
}}
\label{fig:crosssection}
\end{figure}

The third reason is related to the second reason, but concerns
the nature of water Cherenkov detectors. Water Cherenkov
detectors show excellent performance for single ring events or multiple
ring events in which particles forward to opposite directions and the Cherenkov
rings do not overlap with each other. On the other hand, it does
not have good performance for events with heavy ring overlapping.
High-energy ($E_{\nu} \gsim 1~{\rm GeV})$ neutrinos that produce
multiple rings are not preferable in water Cherenkov detectors.

\medskip
The best off-axis angle implies that the peak of the neutrino energy spectrum
is exactly in accordance with the oscillation maximum.
See Table~4 for the relation between the off-axis angle, the peak of the neutrino energy and
the corresponding $\Delta m^{2}_{23}$.
Unfortunately, the $\Delta m^{2}_{23}$ values reported by other experiments\cite{K2K,SKatm} have large
ambiguity, and the best off-axis angle is still unknown.
For this reason, we must maintain the tunability of the off-axis angle until
just before the commissioning of the neutrino beam.

\begin{table}[b!]
\begin{center}
\caption{
Off-axis angle, peak of the neutrino energy spectrum and the corresponding $\Delta m^{2}$ for
a neutrino travel distance of 295~km, which were calculated from $E_{oscmax}=E_{peak}$.}
\smallskip
\begin{tabular}{lcccc}
\hline
\hline
Off axis angle       & ~~~$2.0^{\circ}$~~~ & ~~~$2.1^{\circ}$~~~ & ~~~$2.4^{\circ}$~~~ & ~~~$3.0^{\circ}$~~~ \\
$E_{peak}$ (GeV)     & 0.782   & 0.756  &  0.656  &  0.520  \\
$\Delta m^{2}_{23}(\times 10^{-3}{\rm eV}^{2})$ &  3.28 & 3.17 & 2.75 & 2.18 \\
\hline
\hline
\end{tabular}
\end{center}
\end{table}

We are constructing a decay volume
having a special shape, as shown in Fig.~6.
The cross section of the decay volume is rectangular and
the height of the volume is becoming larger downstream.
The cross section in the most downstream part is
3~m~(width) $\times$ 5.43~m~(height).
If the position of the beam center is adjusted to 1.50~m (3.93~m) below the top
of the decay volume, the direction of the Super-Kamiokande
and Hyper-Kamiokande\cite{T2KLOI,hyper} is
2.0$^{\circ}$(3.0$^{\circ}$) off-axis, as shown in Fig.~6.
The direction of the beam center can be adjusted by arranging the magnets
in the final focusing section and target/horn in the target station.
Civil construction is not required. We can determine the beam direction
about a half year before the beam commissioning which is scheduled for early 2009.
Hopefully, we can decide the exact beam
direction in the summer of 2008, after hearing about the latest result from
the MINOS\cite{MINOS} experiment, reported at the 2008 summer conferences.

\begin{figure}[t!]
\center{{\includegraphics[height=4.0cm]{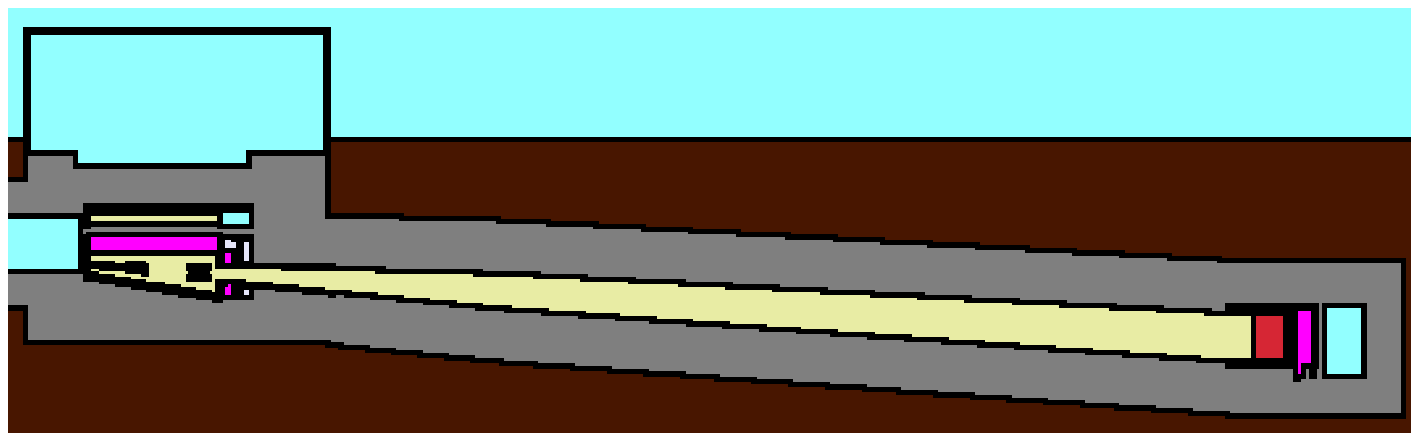}}
\hskip 0.1mm
        {{\includegraphics[height=3.0cm]{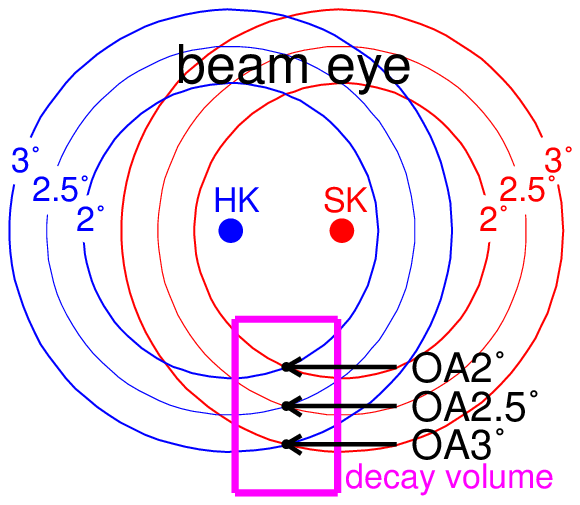}\vskip 1cm}}
}
\center{
\caption{
(left)~Schematic view of the target station, decay volume and beam dump.
The height of the decay volume is becoming larger in the downstream region in order
to change the direction of the beam center.
(right)~Cross section view of the decay volume and the directional correlation
with Super-Kamiokande and Hyper-Kamiokande.
}}
\label{direction}
\end{figure}

\subsection{The 2~km detector}
We have proposed to construct second front detectors at about 2~km from the
proton target for the following two reasons. 

The first reason is related to the off-axis beam.
For reliable neutrino flux extrapolation from the front detectors to the
far detector, the 110~m of the neutrino production point (i.e. decay volume)
must be viewed as \char'134 point like" from the front detectors.
At the 280~m detector with a 2.5$^{\circ}$ off-axis position,
the angular difference between the neutrino from the most upstream part of the
decay volume and the most downstream part is about 1.6$^{\circ}$.
Obviously, this angular difference is not negligible.
The 280~m detector site is
too near to the neutrino production point.

The other reason is the water Cherenkov detector as a front detector.
A water Cherenkov detector is definitely needed as
a front detector because the detection
technique is exactly the same as the far detector, and most of the systematic errors
inherent to the detector can be canceled out.
However, we have a serious difficulty with the event rate.
If a 1~kt water Cherenkov detector
(the same size as K2K) is constructed at the 280~m site,
the event rate will be about 60 events per one spill, which is a 4.2~$\mu$sec beam duration.
This event rate is much larger than the capacity of the water Cherenkov detector.

Because of these reasons, we decided to construct second front detectors,
whose main component is a water Cherenkov detector outside of the JAEA campus.
From physics constraints as well as a problem concerning estate ownership, we proposed them
to be located at about 2~km away from the target. This 2~km detector was not included
in the first proposal, and has not yet been approved. 

\section{Physics goal of T2K}

If neutrinos have mass, the flavor eigenstates are a mixture of the mass eigenstates:
\begin{equation}
\left( \begin{array}{c}\nue \\ \num \\ \nut \end{array} \right) = {\rm{\bf U}}
\left( \begin{array}{c}\nu_{1} \\ \nu_{2} \\ \nu_{3} \end{array} \right).
\end{equation}
The neutrino mass matrix, {\bf U}, has 6 independent parameters. They are
2 square mass differences ($\Delta m^{2}_{12}$ and $\Delta m^{2}_{23}$),
3 mixing angles ($\theta_{12}$, $\theta_{23}$ and $\theta_{13}$)
and 1 CP-violation phase ($\delta$).

Among 6 parameters, $\Delta m^{2}_{12}$ and $\theta_{12}$ were determined
by solar neutrino experiments\cite{SKsolar, SNO}
and a reactor experiment\cite{Kamland}.
At present, their values are
\begin{equation}
\Delta m^{2}_{12} = (6 \sim 8)\times 10^{-5}{\rm eV}^{2}
~~~~{\rm and}~~~\sin^{2}2\theta_{12} \approx 0.8. 
\end{equation}
On the other hand, $\Delta m^{2}_{23}$ and $\theta_{23}$ were studied by
using atmospheric neutrinos\cite{SKatm}
and long-baseline accelerator neutrinos\cite{K2K}.
The present values are
\begin{equation}
\Delta m^{2}_{23} = (2 \sim 3)\times 10^{-3}{\rm eV}^{2}
~~~~{\rm and}~~~\sin^{2}2\theta_{23} \gsim 0.9. 
\end{equation}
Accordingly, unknown oscillation parameters are $\theta_{13}$
and $\delta$.

The physics goal of the T2K experiment is a complete understanding of
the neutrino-oscillation parameters.  It includes:
(1)~the first observation of finite $\theta_{13}$;
(2)~precise measurements of $\Delta m^{2}_{23}$ and $\theta_{23}$;
(3)~observation of the CP violation phase $\delta$
after a beam-intensity upgrade from 0.75 MW to 4MW and
construction of Hyper-Kamiokande. 
The third topic will not occur within the next decade, and is thus beyond the
scope of this document. Only the first and second items are discussed below.

\subsection{\nenm oscillation}

Within the framework of 3-flavor oscillation, the oscillation probability of
$\num$ to $\nue$ with the $\theta_{13}$ channel is written as
\begin{equation}
P(\nu_{\mu} \rightarrow \nu_{e}) \approx \sin^{2}
\theta_{23} \sin^{2}2\theta_{13} \sin^{2}\biggl({{1.27 \Delta m^{2}_{23}L}\over{E_{\nu}}}\biggr).  
\end{equation}
In past neutrino oscillation searches, $\Delta m^{2}$ and
$\sin^{2}2\theta$ were simultaneously searched for in the
$\Delta m^{2}$-$\sin^{2}2\theta$ plane. On the contrary, in the $\theta_{13}$
search, other parameters in the oscillation probability have almost been
determined from the \nmnt oscillation, as given in Eq.(8).

From a theoretical point of view, $\theta_{13}$ is expected to be small because
it is the mixing angle between the first and third generation.
In fact, the present upper limit reported by the CHOOZ experiment\cite{CHOOZ} is
$\sin^{2}2\theta_{13} \sim 0.1$.
Even though the energy of the neutrino beam is adjusted to $E_{oscmax}$ based on
knowledge of the \nmnt oscillation,
\begin{equation}
P(\nu_{\mu} \rightarrow \nu_{e}) \approx {{1}\over{2}} \sin^{2}2\theta_{13}
\sin^{2}\biggl({{1.27 \Delta m^{2}_{23}L}\over{E_{\nu}}}\biggr) < 0.05,
\end{equation}
where $\sin^{2}\theta_{13}$ is taken as $\sim$1/2 from Eq.(8).
Therefore, an appearance search of electron neutrinos with a probability of less than
a few percent is necessary to find a finite $\theta_{13}$.
Obviously, the statistics of K2K was too poor to examine such a small oscillation
probability~(see \cite{K2Kelec} and the first section).

More than 100-times larger statistics in T2K makes a search for such a
small oscillation probability possible. 
Furthermore, in addition to the analysis procedure in K2K,
constraints on the electron neutrino energy can be applied
in selecting the $\nue$ appearance signal,
because the parent neutrino energy calculated from Eq.(5) (by replacing muon by electron)
has a quasi-monochromatic energy spectrum,
as discussed in Section 2.2.

From a careful Monte-Carlo study, about $\sim$100 $\nue$ signals are expected,
whereas the background from the neutral current and the beam-originated $\nue$ are less than 15,
if the oscillation parameter is assumed to be $\sin^{2}2\theta_{13}=0.1$
with $5\times 10^{21}$POT.
The nominal sensitive region is 
\begin{equation}
\sin^{2}2\theta_{13} > 0.006.
\end{equation}
This improves the sensitive region from the CHOOZ experiment by a factor of $\sim$20.

\subsection{\nmnt oscillation}

Precise determinations of $\Delta m^{2}_{23}$ and $\sin^{2}2\theta_{23}$ were attained
by precise measurements of the neutrino survival probability as a function
of the neutrino energy given in Fig.~4.

$\Delta m^{2}_{23}$ is directly determined from the position of
the oscillation maximum ($E_{oscmax}$) from Eq.(3).
From $\sim$5\% accuracy of the $E_{oscmax}$ measurement,
$\Delta m^{2}_{23}$ is also determined with $\sim$5\% accuracy.
Since $\Delta m^{2}_{23} = (2 \sim 3)\times 10^{-3}{\rm eV}^{2}$,
\begin{equation}
\delta(\Delta m^{2}_{23})\sim 0.1\times 10^{-3}{\rm eV^{2}}
\end{equation}
is possible.

On the other hand, $\sin^{2}2\theta_{23}$ is determined
from the depth of the dip at $E_{oscmax}$ in the neutrino survival
probability (see Fig.~4) and/or the absolute reduction rate of
muon neutrino events. Because more than 10000 neutrino events
are expected within 5 years of operation, the statistical error for
the event rate is reduced to be about 1\%. Therefore, 
\begin{equation}
\delta(\sin^{2}2\theta_{23}) \sim 0.01
\end{equation}
is expected.

\bigskip

The author is grateful to the members of K2K and T2K collaborations
for fruitful discussions.


\begin{thebibliography}{99}
\bibitem{hyper} 
    An one mega ton water Cherenkov detector, Hyper-Kamiokande, is under
    preparation for proposal. The location is also in Kamioka area,
    but is about 10km south from the Super-Kamiokande site.
    To provide the same off-axis beam both toward Super-Kamiokande
    and Hyper-Kamiokande,
    the beam center must be moved in vertical.
\end{thebibliography}
\end{document}